\documentclass[12pt, draftclsnofoot, journal, letterpaper, onecolumn]{IEEEtran}

\usepackage{amsfonts}
\usepackage[dvips]{graphicx}
\usepackage{times}
\usepackage{cite}
\usepackage{amsmath}
\usepackage{cases}
\usepackage{array}
\usepackage{dsfont}
\usepackage{amssymb}

\usepackage{stfloats}
\usepackage{slashbox}
\usepackage{graphicx}
\usepackage{footnote}
\usepackage{booktabs}
\usepackage{array}
\usepackage{bm}
\usepackage{multicol}
\usepackage{algorithmic}
\usepackage{algorithm}
\usepackage{empheq}
\usepackage{multirow}
\begin{document}

\title{Optimal DoF Region for the Asymmetric Two-Pair MIMO Two-Way Relay Channel}
\author{\IEEEauthorblockN{Kangqi Liu, \IEEEmembership{Student Member,~IEEE}, Meixia Tao, \IEEEmembership{Senior Member,~IEEE}, and Xiaojun Yuan, \IEEEmembership{Senior Member,~IEEE}}\\
\thanks{Part of this work is submitted to the 2016 IEEE Global Telecommunications Conference. Kangqi Liu and Meixia Tao are with the Department of Electronic Engineering, Shanghai Jiao Tong University, Shanghai, China (Emails: k.liu.cn@ieee.org, mxtao@sjtu.edu.cn). Xiaojun Yuan is with the School of Information Science and Technology, ShanghaiTech University, Shanghai, China (Emails: yuanxj@shanghaitech.edu.cn).}
}

\maketitle

\begin{abstract}
In this paper, we study the optimal degrees of freedom (DoF) region for the two-pair MIMO two-way relay channel (TWRC) with asymmetric antenna setting, where two pairs of users exchange information with the help of a common relay. Each user $i$ is equipped with $M_i$ antennas, for $i=1,2,3,4$, and the relay is equipped with $N$ antennas. First, we derive an outer bound of the DoF region by using the cut-set theorem and the genie-message approach. Then, we propose a new transmission scheme to achieve the outer bound of the DoF region. Due to the asymmetric data exchange, where the two users in each pair can communicate a different number of data streams, we not only need to form the network-coded symbols but also need to process the additional asymmetric data streams at the relay. This is realized through the joint design of relay compression matrix and source precoding matrices. After obtaining the optimal DoF region, we study the optimal sum DoF by solving a linear programming problem. From the optimal DoF region of this channel, we show that in the asymmetric antenna setting, some antennas at certain source nodes are redundant and cannot contribute to enlarge the DoF region. We also show that there is no loss of optimality in terms of the sum DoF by enforcing symmetric data exchange, where the two users in each pair are restricted to communicate the same number of data streams.
\end{abstract}


\section{Introduction}
Due to the great promises in power reduction, coverage extension, and throughput enhancement, wireless relaying has been an important ingredient in both ad hoc and infrastructure-based wireless networks \cite{Hua,Gastpar,Kramer}. Nowadays, a relay has become very much like a wireless gateway where multiple users share a common relay and communicate with each other. A typical representative is the two-way relay channel (TWRC) \cite{Rankov1,Rankov,Wang2}, where two users exchange information with each other through a relay. The success of two-way relaying owes to the invention of physical layer network coding (PLNC) \cite{ZhangSL,Katti}, which almost doubles the spectral efficiency compared with traditional one-way relaying \cite{Vaze,Yuan2}.

A natural generalization of TWRC in multi-user and multi-antenna scenarios is known as multi-user MIMO (MU-MIMO) TWRC \cite{Liu4}, where there are $K$ source nodes and one relay node, all equipped with multiple antennas, and each source node can exchange independent messages with an arbitrary set of other nodes via the relay node. It includes several special cases: $K$-user MIMO Y channel \cite{Lee}, multi-pair MIMO two-way relay channel \cite{Ganesan,Chen,Avestimehr,Sezgin1,Xin}, and $L$-cluster $K$-user MIMO multiway relay channel \cite{Tian,Yuan}. However, the exact capacity analysis for these channels is extremely challenging, only constant-gap capacity is known in the simplest scenario \cite{Chaaban,Sezgin}. As a measure of the approximate capacity in the high signal-to-noise ratio (SNR) region, degrees of freedom (DoF) \cite{Cadambe} specifies how the transmission rate scales as the transmission power goes to infinity. DoF also characterizes the number of interference-free data streams that can be communicated in a given channel.

The DoF analysis for various MU-MIMO TWRC has attracted much attention in the literature \cite{Lee1,Chaaban1,Mu1,Wang3,Wang8,Wang5,Liu4,Tian,Liu5,Yuan,Zewail,Chaaban4}. Such analysis is tractable mainly due to signal alignment proposed in \cite{Lee1} as an integration of PLNC and interference alignment (IA) \cite{Jafar,Maddah}. Recent developments include signal alignment for MIMO Y channel \cite{Lee1}, signal group alignment for $K$-user MIMO Y channel \cite{Mu1}, signal pattern approach for $L$-cluster $K$-user MIMO multiway relay channel \cite{Wang5}, and generalized signal alignment (GSA) for the arbitrary MU-MIMO TWRC \cite{Liu4}. The main results are summarized in \textsc{Table} \ref{table1}. Here, $N$ denotes the number of antennas at the relay node, $M_i$ denotes the number of antennas at each user $i$ for the asymmetric antenna setting, and $M$ denotes the number of antennas at each user for the symmetric antenna setting. It is seen from \textsc{Table} \ref{table1} that the complete characterization of the sum DoF is only available for $K \leq 4$ users with symmetric antenna setting. The analysis of the sum DoF and the DoF region in the general case with asymmetric antenna setting largely remains open.

\begin{table}[]
\tiny
\centering
\caption{Recent Advances towards the DoF Analysis for MU-MIMO TWRC}\label{table1}
\begin{tabular}{|l|l|l|l|l|l|}
\hline
Channel Model                            & Antenna setting        & Sum DoF/DoF region & Antenna configuration for optimal sum DoF/DoF region                                             & Status                & Reference                   \\ \hline
\multirow{2}{*}{MIMO Y channel}          & Symmetric                   & Sum DoF            & $\frac{N}{M} \in (0, +\infty)$                                                & Done                  & \cite{Lee1,Chaaban1}      \\ \cline{2-6}
                                         & Asymmetric                  & Sum DoF            & $(M_1,M_2,M_3,N) \in \mathbb{R}_+^4$                                          & Done                  & \cite{Chaaban1}           \\ \hline
Four-user MIMO Y channel                 & Symmetric                   & Sum DoF            & $\frac{N}{M} \in (0, +\infty)$                                                & Done                  & \cite{Wang8,Liu4}         \\ \hline
\multirow{3}{*}{$K$-user MIMO Y channel} & \multirow{2}{*}{Symmetric}  & Sum DoF            & $\frac{N}{M} \in \big(0, 2+\frac{4}{K(K-1)}\big] \cup \big[K-2, +\infty\big)$ & Partial               & \cite{Liu4}               \\ \cline{3-6}
                                         &                             & DoF region         & $\frac{N}{M} \in (0, 1] \cup [K, + \infty)$                                                      & Partial               & \cite{Chaaban4}           \\ \cline{2-6}
                                         & Asymmetric                  & Sum DoF            & $N \geq \max\{\sum_{i=1}^K M_i-M_s-M_t+d_{s,t}\mid \forall s,t\}$             & Partial               & \cite{Liu5}               \\ \hline
Two-pair MIMO TWRC                       & Symmetric                   & Sum DoF            & $\frac{N}{M} \in (0, +\infty)$                                                & Done                  & \cite{Liu4}               \\ \hline
$\frac{K}{2}$-pair MIMO TWRC             & Symmetric                   & Sum DoF            & $\frac{N}{M} \in \big(0, 2+\frac{4}{K}\big] \cup \big[K-2, +\infty\big)$      & Partial               & \cite{Liu4}               \\ \hline
$L$-cluster $K$-user MIMO multi-way relay channel             & Asymmetric                   & Sum DoF            & Refer to Theorem 2-4 in \cite{Tian}      & Partial               & \cite{Tian}               \\ \hline
\multirow{2}{*}{Two-pair MIMO TWRC}      & \multirow{2}{*}{Asymmetric} & Sum DoF            & \multirow{2}{*}{$(M_1,M_2,M_3,M_4,N) \in \mathbb{R}_+^5$}                     & \multirow{2}{*}{Done} & \multirow{2}{*}{This paper} \\ \cline{3-3}
                                         &                             & DoF region         &                                                                               &                       &                             \\ \hline
\end{tabular}
\end{table}

In this work, we aim to make some progress toward the DoF analysis of the MU-MIMO TWRC with asymmetric antenna setting. To this end, we have succeeded in providing the complete characterization of both DoF region and sum DoF for asymmetric two-pair MIMO TWRC  with antenna configuration $(M_1,M_2,M_3,M_4,N) \in \mathbb{R}_+^5$ for the first time. The main contributions and results of this paper are as follows.

We first derive an outer bound of the DoF region for any antenna configuration by using the cut-set theorem and the genie-message approach. Then we propose a new transmission scheme based on the idea of GSA \cite{Liu4} to achieve the outer bound of the DoF region. Let $d_i$ $(d_{\bar{i}})$ denote the number of interference-free data streams to be transmitted from (to) user $i$ to (from) its pairing user $\bar{i}$. The key idea of the proposed achievable scheme is to align $\min\{d_{i}, d_{\bar{i}}\}$ pairs of bidirectional signals to be exchanged between user $i$ and its pairing user $\bar{i}$ in a same compressed subspace so as to form $\min\{d_{i}, d_{\bar{i}}\}$ network-coded symbols, and project the additional $\max\{d_{i}, d_{\bar{i}}\}-\min\{d_{i}, d_{\bar{i}}\}$ unidirectional data streams from one user to another on a different subspace for complete decoding. This is realized through the joint design of relay compression matrix and source precoding matrices. In \cite{Chaaban1}, the optimal sum DoF of the MIMO Y channel with asymmetric antenna setting is characterized by using signal alignment and antenna deactivation techniques. It is pointed out that symmetric data exchange, where the two users in each pair communicate the same number of data streams, can achieve the optimal sum DoF. In \cite{Liu5}, the sum DoF of an arbitrary MU-MIMO TWRC is analyzed under an asymmetric antenna setting. But the analysis is limited to symmetric data exchange. In contrast with the scheme in \cite{Chaaban1} and \cite{Liu5}, we need not only to construct network-coded symbols but also to process the additional asymmetric data streams at the relay, which enables us to obtain the DoF region rather than just the sum DoF. In \cite{Chaaban4}, the optimal DoF region of the $K$-user MIMO Y channel with symmetric antenna setting is studied by using channel diagonalization and cyclic communication techniques. Their transmission scheme is only applicable when the antenna configuration satisfies $\frac{N}{M} \in (0, 1] \cup [K, + \infty)$. By comparing with the scheme in \cite{Chaaban4}, our transmission scheme is applicable for all different antenna configurations.

After obtaining the optimal DoF region, determining the optimal sum DoF becomes a linear programming problem. By analyzing this problem, we find that enforcing symmetric data exchange within each user pair does not lose any optimality in terms of the sum DoF. Based on this finding, the linear programming problem is greatly simplified and we are able to obtain the optimal sum DoF explicitly at all antenna configurations.

The rest of the paper is organized as follows. In Section II, we present the system model. In Section III, we introduce the main results and show the insights of the results. The proof of DoF-region converse and DoF-region achievability are presented in Section IV and Section V, respectively. In Section VI, we show the optimal sum DoF of the channel. Finally, we conclude the paper in Section VII.

Notations: Scalars, vectors, and matrices are denoted by lowercase regular letters, lowercase bold letters, and uppercase bold letters, respectively. $(\cdot)^{T}$ and $(\cdot)^{H}$ denote the transpose and the Hermitian transpose, respectively. rank$({\bf X})$ stands for the rank of ${\bf X}$. {\bf I} is the identity matrix. $\textrm{span} ({\bf X})$ and ${\textrm{null} ({\bf X})}$ stand for the column space and the null space of the matrix ${\bf X}$, respectively. $\binom{n}{m}=\frac{n!}{m!(n-m)!}$ denotes the binomial coefficient indexed by $n$ and $m$.

\section{Channel Model}
\begin{figure}[t]
\begin{centering}
\includegraphics[scale=0.46]{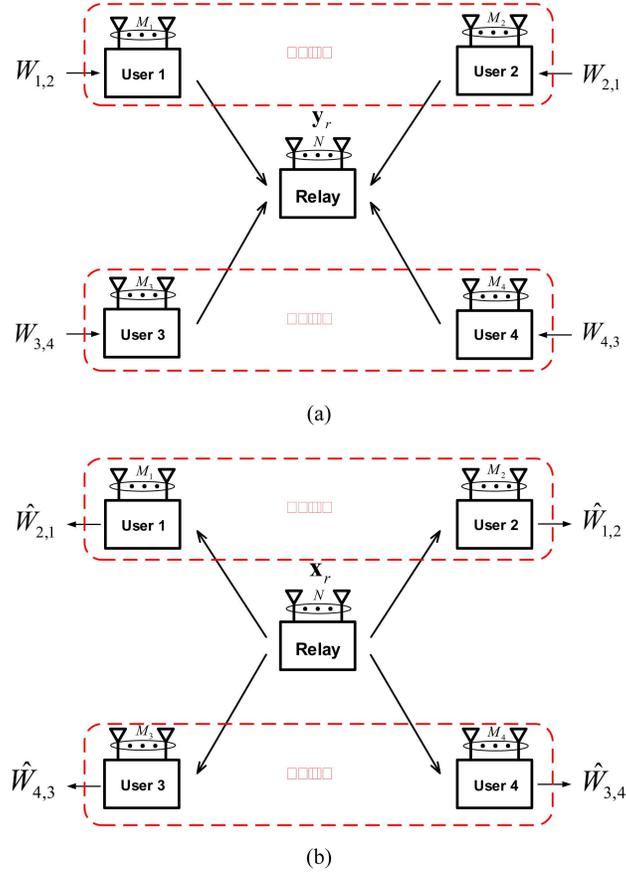}
\vspace{-0.1cm}
 \caption{Asymmetric two-pair MIMO TWRC. (a) Multiple access phase. (b) Broadcast phase.}\label{two_pair_MAC_BC}
\end{centering}
\vspace{-0.3cm}
\end{figure}

Consider a discrete memoryless asymmetric two-pair MIMO TWRC as shown in Fig. \ref{two_pair_MAC_BC}, where users $1$ and $2$ form a pair to exchange information and users $3$ and $4$ form another pair to exchange information, both with the help of a common relay. Each user $i$ $(i=1,2,3,4)$ is equipped with $M_i$ antennas, and the relay is equipped with $N$ antennas. Without loss of generality, we assume that
\begin{equation}
\left\{
  \begin{array}{ll}
    M_1\geq M_2 \\
    M_3 \geq M_4 \\
    M_1+M_2 \geq M_3+M_4.
  \end{array}
\right.
\end{equation}
Denote by ${\bf H}_{i,r}(t) \in \mathbb{C}^{N \times M_i}$ the channel matrix from user $i$ to the relay for channel use $t$, and by ${\bf H}_{r,i}(t) \in \mathbb{C}^{M_i \times N}$ the channel matrix from the relay to user $i$. It is assumed that the entries of the channel matrices are drawn independently from a continuous distribution, which guarantees that the channel matrices have full rank with probability one. Perfect channel knowledge is assumed to be available at each node, and all the nodes in the network are assumed to be full duplex. The message transmitted from user $i$ to its pairing user $\bar{i}$ is denoted by $W_{i,\bar{i}}$, and is independent of each other. Each $W_{i,\bar{i}}$ is encoded using a codebook with size $2^{nR_{i}}$, where $n$ is the codeword length and $R_{i}$ is the information rate of $W_{i,\bar{i}}$. Note that $R_i$ can be different from $R_{i'}$ due to the asymmetric antenna setting, different channel gain, or different rate requirement on user $i$ and $i'$.

The message exchange takes place in two phases: the multiple access (MAC) phase and the broadcast (BC) phase. In the MAC phase, all the users transmit their signals to the relay. The received signal for the channel use $t$ at the relay, denoted by ${\bf y}_{r}(t) \in \mathbb{C}^{N \times 1}$, is given by
\begin{equation}\label{y_r}
{\bf y}_{r}(t)=\sum\limits_{i=1}^4 {\bf H}_{i,r}(t){\bf x}_{i}(t)+{\bf n}_r(t),
\end{equation}
where ${\bf x}_{i}(t) \in \mathbb{C}^{M_i \times 1}$ denotes transmitted signal from user $i$ with average power constraint $\mathds{E}[{\bf x}_{i}(t)^H{\bf x}_{i}(t)]\leq P$, and ${\bf n}_r(t) \in \mathbb{C}^{N \times 1}$ denotes the additive white Gaussian noise (AWGN) vector for the channel use $t$ with each element being independent, and having zero mean and unit variance.

Upon receiving ${\bf y}_r(t)$ in \eqref{y_r}, the relay processes these messages to obtain a mixed signal ${\bf x}_{r}(t) \in \mathbb{C}^{N \times 1}$ with average power constraint $\mathds{E}[{\bf x}_{i}(t)^H{\bf x}_{i}(t)]\leq P$, and broadcasts to all the users. The received signal for the channel use $t$ at user $i$, denoted by ${\bf y}_{i}(t) \in \mathbb{C}^{M_i \times 1}$, is given by
\begin{eqnarray}\label{y_i}
{\bf y}_i(t)={\bf H}_{r,i}(t){\bf x}_r(t)+{\bf n}_i(t),
\end{eqnarray}
where ${\bf n}_i(t) \in \mathbb{C}^{M_i \times 1}$ denotes the AWGN vector for the channel use $t$ with each element being independent, and having zero mean and unit variance.

Each user $i$ will decode its desired message, denoted by $\hat{W}_{\bar{i},i}$, based on the received signals $\{{\bf y}_i(t)\}_{t=1}^n$ and its own transmitted message. Let $R_{i}(P)$ denote the achievable information rate of the message $W_{i,\bar{i}}$ under the power constraint $P$. Here, we say that a rate tuple $\{R_{i}(P)\}_{i=1}^4$ is achievable if
\begin{equation}
\lim\limits_{n\rightarrow \infty} \textrm{Pr}\left(\hat{W}_{i,\bar{i}} \neq W_{i,\bar{i}}\right)=0,~\forall i.
\end{equation}
The DoF of message $W_{i,\bar{i}}$ is defined as
\begin{equation}\label{dof_message}
d_{i} \triangleq \lim\limits_{P \rightarrow \infty} \frac{R_{i}(P)}{\log(P)}.
\end{equation}
The sum DoF is defined as
\begin{equation}\label{sum_dof}
d_{\Sigma}=\sum\limits_{i=1}^4 d_{i}.
\end{equation}
The DoF region is defined as \cite{Jafar}
\begin{align}\nonumber
{\cal D}=\Big\{&(d_1,d_2,d_3,d_4) \in \mathbb{R}_{+}^6: \forall(\omega_{1},\omega_{2},\omega_{3},\omega_{4}) \in \mathbb{R}_{+}^6\\
&\sum\limits_{i=1}^4 \omega_{i}d_{i} \leq \limsup_{P \rightarrow \infty}\left[\sup_{R(P)\in {\cal C}(P)} \left[\sum\limits_{i=1}^4 \omega_{i}R_{i}(P)\right]\frac{1}{\log(P)}\right]\Big\},
\end{align}
where ${\cal C}(P)$ is the capacity region of the asymmetric two-pair MIMO TWRC, which is the set of all achievable rate tuples $\{R_{i}(P)\}_{i=1}^4$. The goal of this work is to characterize the optimal DoF region, denoted by ${\cal D}^{*}$, as well as the optimal sum DoF, denoted by $d_{\Sigma}^{*}$, for the considered asymmetric two-pair MIMO TWRC with antenna configuration $(M_1, M_2, M_3, M_4, N)$.

\section{Main Results}
The main findings of this paper are summarized in the following theorem and corollary.

\textit{Theorem 1}: For the asymmetric two-pair MIMO TWRC with antenna configuration $(M_1, M_2,$ $M_3, M_4, N)$, the optimal DoF region can be expressed as

\begin{subequations}\label{D}
\begin{align}\nonumber
{\cal D}^{*}=&\Big\{(d_1,d_2,d_3,d_4) \in \mathbb{R}_{+}^4:\\\label{D_a}
&~~~d_{1} \leq M_2 \\\label{D_b}
&~~~d_{2} \leq M_2 \\\label{D_c}
&~~~d_{3} \leq M_4 \\\label{D_d}
&~~~d_{4} \leq M_4 \\\label{D_e}
&~~~d_{1}+d_{3} \leq N \\\label{D_f}
&~~~d_{1}+d_{4} \leq N \\\label{D_g}
&~~~d_{2}+d_{3} \leq N \\\label{D_h}
&~~~d_{2}+d_{4} \leq N \\\label{D_i}
&~~~d_{1}+d_{2}+d_{3} \leq \max\{M_1+M_2,N\} \\\label{D_j}
&~~~d_{1}+d_{2}+d_{4} \leq \max\{M_1+M_2,N\} \\\label{D_k}
&~~~d_{1}+d_{3}+d_{4} \leq \max\{M_3+M_4,N\} \\\label{D_l}
&~~~d_{2}+d_{3}+d_{4} \leq \max\{M_3+M_4,N\} \Big\}.
\end{align}
\end{subequations}

\textit{Corollary 1}: For the asymmetric two-pair MIMO TWRC with antenna configuration $(M_1, M_2,$ $M_3, M_4, N)$, the optimal sum DoF is given as follows:
\begin{enumerate}
  \item When $N \geq M_1+M_2$,
  \begin{equation}
  d_{\Sigma}^{*}=\min\left\{2M_2+2M_4, \frac{4}{3}N, M_2+N, M_4+N\right\};
  \end{equation}
  \item When $M_3+M_4 \leq N < M_1+M_2$,
  \begin{equation}
  d_{\Sigma}^{*}=\min\left\{2M_2+2M_4, M_2+N, M_1+M_2+M_4, 2N, \frac{2(M_1+M_2+N)}{3}\right\};
  \end{equation}
  \item When $N < M_3+M_4$,
  \begin{equation}
  d_{\Sigma}^{*}=\min\left\{2M_2+2M_4, 2N, M_2+M_3+M_4, M_1+M_2+M_4, \frac{2(M_1+M_2+M_3+M_4)}{3}\right\}.
  \end{equation}
\end{enumerate}

The DoF converse of \textit{Theorem 1} is proved in Section IV via the cut-set theorem and the genie-message approach. The achievability of \textit{Theorem 1} is proved in Section V. The proof of \textit{Corollary 1} is presented in Section VI.

\textit{Remark 1} (\textit{Redundant antennas}): It is observed from \textit{Theorem 1} that the DoF only depends on $\{M_2, M_4, N\}$ and does not depend on $\{M_1, M_3\}$ when $N \geq M_1+M_2$. This means that if the relay antenna number is large enough, the smaller antenna number within each user pair limits the DoF. Hence, there are $M_1-M_2$ redundant antennas at user $1$, and $M_3-M_4$ redundant antennas at user $3$. Likewise, when $M_3+M_4 \leq N < M_1+M_2$, the DoF only depends on $\{M_1, M_2, M_4, N\}$ and does not depend on $M_3$. Hence, there are $M_3-M_4$ redundant antennas at user $3$.

\textit{Remark 2}: (\textit{Connection to symmetric two-pair MIMO TWRC}): When $M_i=M$, for $i=1,2,3,4$, the sum DoF characterized in \textit{Corollary 1} reduces to $\min\{4M,\max\{\frac{4N}{3},\frac{8M}{3}\},2N\}$, which is consistent with the results in \cite{Liu4}.

\textit{Remark 3}: (\textit{Comparison to the existing work \cite{Tian}}): The authors in \cite{Tian} study the sum DoF for the asymmetric $L$-cluster $K$-user MIMO multi-way relay channel. In the special case when $L=K=2$, the channel in \cite{Tian} reduces to our considered two-pair MIMO TWRC. However, the maximum sum DoF results in \cite{Tian} are neither optimal nor complete, while our sum DoF results in \textit{Corollary 1} are optimal and complete.

\section{DoF-region converse}
The first four bounds in \eqref{D} can be proved easily from the cut-set theorem \cite{Cover}. That is, since each user $i$ has $M_i$ antennas only, the DoF of the transmitted or received message for user $i$ cannot be greater than $M_i$.

We now prove the bound \eqref{D_e} by using the genie-aided message approach as in \cite{Chaaban1,Wang8,Wang3,Liu4}. By the converse assumption, each user $i$ can decode its intended message $\{W_{\bar{i},i}\}$ with its own transmitted messages $\{W_{i,\bar{i}}\}$ as side information. Given the fact that the signal received by each user is a degraded version of the signal received at the relay, if a genie provides the side information $W_{i,\bar{i}}$ to the relay, then the relay is able to decode $W_{\bar{i},i}$. As such, we provide ${\cal G}_1=\{W_{2,1},W_{4,3}\}$ as the genie message to the relay and obtain the following bound:
\begin{subequations}\label{genie_Y_3}
\begin{align}\nonumber
&n(R_{1}+R_{3}-\epsilon)\\\label{genie_Y_3_a}
\leq& I(W_{1,2};{\bf y}_2^n \mid W_{2,1})+I(W_{3,4};{\bf y}_4^n \mid W_{4,3})\\\label{genie_Y_3_b}
\leq& I(W_{1,2};{\bf y}_r^n \mid W_{2,1})+I(W_{3,4};{\bf y}_r^n \mid W_{4,3})\\\label{genie_Y_3_c}
\leq& I(W_{1,2};{\bf y}_r^n \mid {\cal G}_1)+I(W_{3,4};{\bf y}_r^n \mid {\cal G}_1)\\\label{genie_Y_3_d}
\leq & I(W_{1,2},W_{3,4};{\bf y}_r^n \mid {\cal G}_1)\\\label{genie_Y_3_e}
\leq & h({\bf y}_r^n \mid {\cal G}_1)\\\label{genie_Y_3_f}
\leq&  nN\log P,
\end{align}
\end{subequations}
where \eqref{genie_Y_3_a} follows from the Fano's inequality; \eqref{genie_Y_3_b} follows from the data processing inequality; \eqref{genie_Y_3_c} follows from the fact that $I(A;B \mid C,D) \geq I(A;B\mid C)$ when $A$ is independent of $D$; \eqref{genie_Y_3_d} follows from the chain rule. Dividing $n\log P$ through both sides of \eqref{genie_Y_3} and letting $n\rightarrow \infty$ and $P \rightarrow \infty$, we obtain the bound \eqref{D_e}. Similarly, \eqref{D_f}-\eqref{D_h} can be obtained.

Next, we prove the bound \eqref{D_i} through the genie-message approach. Note that there are in total four messages received at the relay. If the message $W_{3,4}$ and $W_{4,3}$ are known at the relay, then the relay can decode $\{W_{1,2},W_{2,1}\}$ provided $N \geq M_{1}+M_{2}$. Hence, we provide $\{W_{4,3}\}$ as a genie message to the relay in the case of $N \geq M_{1}+M_{2}$ in the first step. By the converse assumption, we can obtain the following bound:
\begin{subequations}\label{genie_Y_4}
\begin{align}\nonumber
&n(R_{3}-\epsilon)\\\label{genie_Y_4_a}
\leq& I(W_{3,4};{\bf y}_4^n \mid W_{4,3})\\\label{genie_Y_4_b}
\leq& I(W_{3,4};{\bf y}_r^n \mid W_{4,3})\\\label{genie_Y_4_c}
\leq & h({\bf y}_r^n \mid W_{4,3})-h({\bf y}_r^n \mid W_{3,4},W_{4,3})\\\label{genie_Y_4_d}
\leq&  h({\bf y}_r^n)-h(W_{1,2},W_{2,1},{\bf n}_r^n)\\\label{genie_Y_4_e}
\leq& nN\log P-n(R_1+R_2),
\end{align}
\end{subequations}
Dividing $n\log P$ through both sides of \eqref{genie_Y_4} and letting $n\rightarrow \infty$ and $P \rightarrow \infty$, we obtain
\begin{align}\label{genie_Y_aaa}
d_1+d_2+d_3\leq N
\end{align}
when $N \geq M_{1}+M_{2}$.

\begin{figure}[t]
\begin{centering}
\includegraphics[scale=0.6]{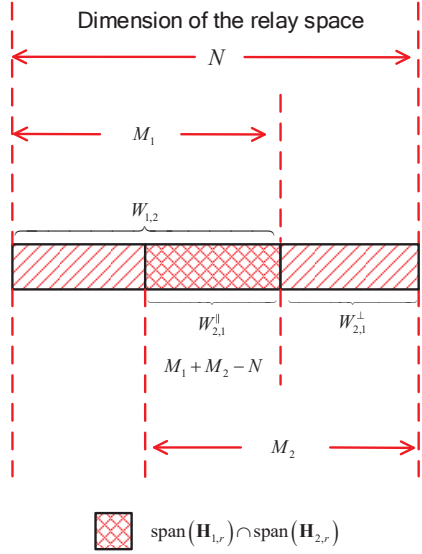}
\vspace{-0.1cm}
 \caption{$W_{1,2}$ and $W_{2,1}$ at the relay.}\label{genie_message_12}
\end{centering}
\vspace{-0.3cm}
\end{figure}

What remains is to consider the case of $N < M_{1}+M_{2}$. Again, if the messages, $W_{3,4}$ and $W_{4,3}$, are already known at the relay, the unknown messages at the relay remain $\{W_{1,2},W_{2,1}\}$, which is illustrated in Fig. \ref{genie_message_12}. It can be seen that there is an intersection subspace of $\textrm{span}({\bf H}_{1,r})$ and $\textrm{span}({\bf H}_{2,r})$ with dimension $M_{1}+M_{2}-N$. We separate the message $W_{2,1}$ into two parts as $W_{2,1}^{\|}$ and $W_{2,1}^{\perp}$, where $W_{2,1}^{\|}$ is located in the intersection subspace and $W_{2,1}^{\perp}$ is orthogonal to the intersection subspace. Then, if $W_{2,1}^{\|}$ is provided as a genie message to the relay, the relay can decode $W_{1,2}$ and $W_{2,1}$ surely. Hence, we provide $\{W_{4,3},W_{2,1}^{\|}\}$ as a genie message to the relay. By the converse assumption, we obtain the following bound:
\begin{subequations}\label{genie_Y_5}
\begin{align}\nonumber
&n(R_{3}-\epsilon)\\\label{genie_Y_5_a}
\leq& I(W_{3,4};{\bf y}_4^n \mid W_{4,3})\\\label{genie_Y_5_b}
\leq& I(W_{3,4};{\bf y}_r^n \mid W_{4,3})\\\label{genie_Y_5_c}
\leq & I(W_{3,4};{\bf y}_r^n,W_{2,1}^{\|} \mid W_{4,3})\\\label{genie_Y_5_d}
\leq & h({\bf y}_r^n,W_{2,1}^{\|} \mid W_{4,3})-h({\bf y}_r^n,W_{2,1}^{\|} \mid W_{3,4},W_{4,3})\\\label{genie_Y_5_e}
\leq&  h({\bf y}_r^n)+h(W_{2,1}^{\|})-h(W_{1,2},W_{2,1},{\bf n}_r^n)\\\label{genie_Y_5_f}
\leq& nN\log P+n(M_1+M_2-N)\log P-n(R_1+R_2).
\end{align}
\end{subequations}
Dividing $n\log P$ through both sides of \eqref{genie_Y_5} and letting $n\rightarrow \infty$ and $P \rightarrow \infty$, we obtain
\begin{align}\label{genie_Y_bbb}
d_1+d_2+d_3\leq M_{1}+M_{2}
\end{align}
when $N < M_{1}+M_{2}$.
Combining \eqref{genie_Y_aaa} and \eqref{genie_Y_bbb}, we obtain the bound \eqref{D_i}. Similarly, \eqref{D_j}-\eqref{D_l} hold, which concludes the proof.

\section{DoF-region achievability}
In this section, we prove the achievability of the optimal DoF region for the asymmetric two-pair MIMO TWRC. We first illustrate the main idea of our proposed transmission scheme using an example. Then we consider the general case and present the achievable schemes in three different antenna configurations: (I) $N \geq M_1+M_2$; (II) $M_3+M_4 \leq N < M_1+M_2$; (III) $N < M_3+M_4$.

\subsection{An example with $(M_1, M_2, M_3, M_4, N)=(6,5,4,4,9)$}
In this subsection, we illustrate how to achieve the DoF tuple ${\bf d}=(5,3,3,1)$ under the antenna configuration $(M_1, M_2, M_3, M_4, N)=(6,5,4,4,9)$.  In this example, there are $\min\{5,3\}=3$ pairs of data streams to be exchanged between user $1$ and $2$, and $\min\{3,1\}=1$ pair of data streams to be exchanged between user $3$ and user $4$. In addition to that, user $1$ has $2$ more data streams to communicate with user $2$ and user $3$ has $2$ more data streams for user $4$.

\begin{figure}[t]
\begin{centering}
\includegraphics[scale=0.7]{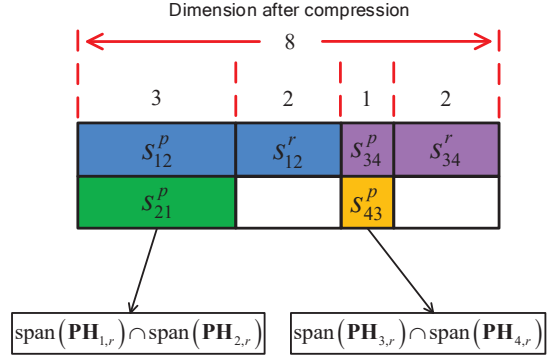}
\vspace{-0.1cm}
 \caption{Alignment in the MAC phase.}\label{Example_2}
\end{centering}
\vspace{-0.3cm}
\end{figure}

During the MAC phase, the signal received at the relay can be rewritten as
\begin{align}\label{y_r_E}
{\bf y}_r=\sum\limits_{i=1}^4 {\bf H}_{i,r}{\bf V}_i^p{\bf s}_i^p+{\bf H}_{1,r}{\bf V}_{1}^r{\bf s}_{1}^r+{\bf H}_{3,r}{\bf V}_{3}^r{\bf s}_{3}^r+{\bf n}_r.
\end{align}
Here, ${\bf s}_1^p \in {\mathbb C}^{3 \times 1}$ and ${\bf s}_2^p \in {\mathbb C}^{3 \times 1}$ are the pair of signals to be exchanged between user $1$ and user $2$, ${\bf s}_3^p \in {\mathbb C}$ and ${\bf s}_4^p \in {\mathbb C}$ are the pair of signals to be exchanged between user $3$ and user $4$, ${\bf s}_1^r \in {\mathbb C}^{2 \times 1}$ and ${\bf s}_3^r \in {\mathbb C}^{2 \times 1}$ represent the additional signals sent from user $1$ and user $3$ to user $2$ and user $4$, respectively; ${\bf V}_{1}^p \in {\mathbb C}^{6 \times 3}$, ${\bf V}_{2}^p \in {\mathbb C}^{5 \times 3}$, ${\bf V}_{3}^p \in {\mathbb C}^{4 \times 1}$, ${\bf V}_{4}^p \in {\mathbb C}^{4 \times 1}$, ${\bf V}_{1}^r \in {\mathbb C}^{6 \times 2}$, and ${\bf V}_{3}^r \in {\mathbb C}^{4 \times 2}$ are the corresponding precoding matrices. According to the GSA principle proposed in \cite{Liu4}\footnote{GSA refers to that a pair of signals to be exchanged are aligned at a same compressed subspace at the relay through the joint design of relay compression matrix and source precoding matrices.}, we need to jointly design a full-rank relay compression matrix ${\bf P} \in {\mathbb C}^{J \times 9}$ and all the precoding matrices $\{{\bf V}_{i}^p \mid i=1,2,3,4\}$ and $\{{\bf V}_{i}^r \mid i=1,3\}$ such that:
\begin{subequations}\label{GSA_E}
\begin{align}\label{GSA_E_1}
&{\bf P}{\bf H}_{1,r}{\bf V}_1^p={\bf P}{\bf H}_{2,r}{\bf V}_2^p,\\\label{GSA_E_2}
&{\bf P}{\bf H}_{3,r}{\bf V}_3^p={\bf P}{\bf H}_{4,r}{\bf V}_4^p,\\\label{GSA_E_3}
&\textrm{rank}([{\bf V}_{1}^p~{\bf V}_{1}^r])=5,\\\label{GSA_E_4}
&\textrm{rank}([{\bf V}_{3}^p~{\bf V}_{3}^r])=3.
\end{align}
\end{subequations}
A signal space illustration is given in Fig. \ref{Example_2}. Specifically, condition \eqref{GSA_E_1} means that the relay needs to align the signal pair $({\bf s}_1^p, {\bf s}_2^p)$ in a subspace to form network-coded symbols, and condition \eqref{GSA_E_2} means to align the signal pair $({\bf s}_3^p, {\bf s}_4^p)$ in another subspace to form network-coded symbols. Condition \eqref{GSA_E_3} is to ensure the separability of ${\bf s}_1^p$ and ${\bf s}_1^r$ at user $1$, and likewise condition \eqref{GSA_E_4} is to ensure the separability of ${\bf s}_3^p$ and ${\bf s}_3^r$ at user $3$. In total, the relay needs to decode $8$ independent symbols and we should choose $J=8$ according to \cite{Liu4}.

However, in \cite{Liu4}, the authors only provide the necessary and sufficient condition for the GSA equation to hold under the symmetric antenna setting when $N \geq 2M$. In the following lemma, we give the necessary and sufficient condition for \eqref{GSA_E_1} and \eqref{GSA_E_2} to hold under the general asymmetric antenna setting.

\textit{Lemma 1}: The GSA equations \eqref{GSA_E_1} and \eqref{GSA_E_2} hold if and only if there are at least $J-M_i-M_{\bar {i}}+d_{i}$ basis vectors of $\textrm{span}\left({\bf P}^T\right)$ that lie in the null space of $\left[{{\bf H}}_{i,r}~~-{{\bf H}}_{\bar {i},r}\right]^T$ for all user pair $(i, \bar {i})$ with $M_i+M_{\bar {i}}-J < d_{i}$.
\begin{proof}
The proof is similar to the proof of \textit{Theorem 4} in \cite{Liu4} and thus omitted here.
\end{proof}

It is noted that the dimension of the intersection space between ${\bf P}{\bf H}_{1,r}$ and ${\bf P}{\bf H}_{2,r}$ is 3, which is enough to align the data streams ${\bf s}_1^p$ and ${\bf s}_2^p$. The difficulty is how to align ${\bf s}_3^p$ and ${\bf s}_4^p$ as there is no intersection subspace between ${\bf P}{\bf H}_{3,r}$ and ${\bf P}{\bf H}_{4,r}$. To this end, we first design ${\bf P}$, ${\bf V}_3^p$ and ${\bf V}_4^p$ for \eqref{GSA_E_2} to hold. According to \textit{Lemma 1}, there should be one row of $\bf P$ that lies in the left null space $[{\bf H}_{3,r}~-{\bf H}_{4,r}]$ for \eqref{GSA_E_2} to hold. Thus, we design $\bf P$ such that
\begin{align}
{\bf P}=\left[\begin{array}{c}
                {\bf P}_1 \\
                {\bf P}_2
              \end{array}
\right],
\end{align}
where ${\bf P}_1$ is a $1 \times 9$ submatrix satisfying
\begin{align}\label{I_A_PP_E_1}
{\bf P}_1[{\bf H}_{3,r}~-{\bf H}_{4,r}]=0,
\end{align}
or equivalently,
\begin{align}\label{I_A_PP_E_2}
&\textrm{span}\left({\bf P}_1^T\right)\subseteq\textrm{null} \left(\left[{\bf H}_{3,r}~-{\bf H}_{4,r}\right]\right)^T
\end{align}
and ${\bf P}_2$ is a $7 \times 9$ submatrix that can be designed randomly as long as ${\bf P}$ has full row rank. For instance, we can choose ${\bf P}_2=[{\bf I}~{\bf 0}]$. Then, ${\bf P}\left[{\bf H}_{3,r}~-{\bf H}_{4,r}\right]$ can be expressed as
\begin{align}
{\bf P}\left[{\bf H}_{3,r}~-{\bf H}_{4,r}\right]=\left[\begin{array}{cc}
                                                 \bf 0 & \bf 0 \\
                                                 {\bf P}_2{\bf H}_{3,r} & {\bf P}_2{\bf H}_{4,r}
                                               \end{array}
\right],
\end{align}
with rank $7$. We then design ${\bf V}_3^p$ and ${\bf V}_4^p$ as
\begin{align}\label{I_A_V_2_E}
\textrm{span}\left(\left[\begin{array}{c}
        {\bf V}_3^p \\
        {\bf V}_4^p
      \end{array}
\right] \right) \subseteq \textrm{null} \left(\left[{\bf P}_2{\bf H}_{3,r}~~{\bf P}_2{\bf H}_{4,r}\right]\right).
\end{align}
Here, ${\bf V}_3^p$ and ${\bf V}_4^p$ exist because the dimension of the null space of ${\bf P}_2\left[{\bf H}_{3,r}~-{\bf H}_{4,r}\right]$ is $1$.

Next, we design ${\bf V}_1^p$ and ${\bf V}_2^p$ for \eqref{GSA_E_1} to hold. It is noted that the rank of ${\bf P}\left[{\bf H}_{1,r}~-{\bf H}_{2,r}\right]$ is 8 and hence the null space of the $8 \times 11$ matrix ${\bf P}\left[{\bf H}_{1,r}~-{\bf H}_{2,r}\right]$ has dimension of 3. We can design ${\bf V}_1^p$ and ${\bf V}_2^p$ as
\begin{align}\label{I_A_V_1_E}
\textrm{span}\left(\left[\begin{array}{c}
        {\bf V}_1^p \\
        {\bf V}_2^p
      \end{array}
\right] \right) \subseteq \textrm{null} \left({\bf P}\left[{\bf H}_{1,r}~-{\bf H}_{2,r}\right]\right).
\end{align}

Once $\{{\bf V}_i^p\mid i=1,2,3,4\}$ are designed, ${\bf V}_1^r$ (or ${\bf V}_3^r$) can be designed randomly as long as $[{\bf V}_1^p~{\bf V}_1^r]$ and $[{\bf V}_3^p~{\bf V}_3^r]$ have full column rank in order to meet \eqref{GSA_E_3} and \eqref{GSA_E_4}. The signal after compression at the relay can be expressed as
\begin{align}\nonumber
{\bf P}{\bf y}_r=&{\bf P}{\bf H}_{1,r}{\bf V}_1^p({\bf s}_1^p+{\bf s}_2^p)+{\bf P}{\bf H}_{3,r}{\bf V}_3^p({\bf s}_3^p+{\bf s}_4^p)\\\label{Pyr1}
&+{\bf P}{\bf H}_{1,r}{\bf V}_{1}^r{\bf s}_{1}^r+{\bf P}{\bf H}_{3,r}{\bf V}_{3}^r{\bf s}_{3}^r+{\bf n}_r.
\end{align}

Thus far, the relay is able to decode the network-coded symbols, ${\bf s}_1^p+{\bf s}_2^p$ and ${\bf s}_3^p+{\bf s}_4^p$, as well as the remaining symbols, ${\bf s}_{1}^r$ and ${\bf s}_{3}^r$, by using an $8 \times 8$ zero-forcing matrix
\begin{align}\label{W}
{\bf W}=\left([{\bf P}{\bf H}_{1,r}{\bf V}_1~~{\bf P}{\bf H}_{3,r}{\bf V}_3~~{\bf P}{\bf H}_{1,r}{\bf V}_{1}^r~~{\bf P}{\bf H}_{3,r}{\bf V}_{3}^r]\right)^{-1}.
\end{align}
The decoded symbol vector, $\hat{{\bf s}}_r \in {\mathbb C}^{8 \times 1}$, can be expressed as
\begin{align}\label{s_r}
\hat{{\bf s}}_r&={\bf W}{\bf P}{\bf y}_r=\left[\begin{array}{c}
                                                                                                     {\bf s}_1^p+{\bf s}_2^p \\
                                                                                                     {\bf s}_3^p+{\bf s}_4^p \\
                                                                                                     {\bf s}_1^r\\
                                                                                                     {\bf s}_3^r
                                                                                                   \end{array}
\right]+{\bf W}{\bf P}{\bf n}_r\\
&={\bf s}_r+{\bf W}{\bf P}{\bf n}_r.
\end{align}

We next introduce the transmission scheme in the BC phase for each user to decode its desired message. The signal received at user $i$ with receiving matrix ${\bf U}_{i} \in {\mathbb C}^{d_{\bar{i}} \times M_i}$ can be expressed as
\begin{align}\nonumber
\hat{{\bf s}}_i=&{\bf U}_i{\bf y}_i+{\bf U}_i{\bf n}_i\\\label{s_i_E}
=&{\bf U}_i{\bf H}_{r,i}{\bf Q}{\bf T}{\bf s}_r+{\bf U}_i{\bf H}_{r,i}{\bf Q}{\bf T}{\bf W}{\bf P}{\bf n}_r+{\bf U}_i{\bf n}_i,
\end{align}
where ${\bf Q} \in {\mathbb C}^{9\times 8}$ denotes a compression matrix in the BC phase and ${\bf T} \in {\mathbb C}^{8 \times 8}$ denotes a zero-forcing matrix in the BC phase.

Due to the symmetry between the MAC and BC phases, we redefine ${\bf U}_i$ as
\begin{align}
&{\bf U}_i={\bf U}_i^p,~i=1,3,\\
&{\bf U}_i=\left[
            \begin{array}{c}
              {\bf U}_i^p \\
              {\bf U}_i^r \\
            \end{array}
          \right],~i=2,4
\end{align}
satisfying
\begin{subequations}\label{GSA_BC}
\begin{align}\label{GSA_BC_1}
&{\bf U}_1^p{\bf H}_{r,1}{\bf Q}={\bf U}_2^p{\bf H}_{r,2}{\bf Q}\\\label{GSA_BC_2}
&{\bf U}_3^p{\bf H}_{r,3}{\bf Q}={\bf U}_4^p{\bf H}_{r,4}{\bf Q}.
\end{align}
\end{subequations}
Note that there exists symmetry between the design of $\bf P$ and $\bf Q$, ${\bf V}_i^p$ and ${\bf U}_i^p$, as well as ${\bf V}_i^r$ and ${\bf U}_{\bar{i}}^r$. Then the zero-forcing matrix in the BC phase ${\bf T}$ can be designed as
\begin{align}\label{T}
{\bf T}=\left(\left[\begin{array}{c}
                      {\bf U}_2^p{\bf H}_{r,2}{\bf Q} \\
                      {\bf U}_4^p{\bf H}_{r,4}{\bf Q} \\
                      {\bf U}_2^r{\bf H}_{r,2}{\bf Q} \\
                      {\bf U}_4^r{\bf H}_{r,4}{\bf Q}
                    \end{array}
\right]\right)^{-1}.
\end{align}
The signal received at user $i$ in \eqref{s_i_E} can be rewritten as
\begin{align}\label{s_i_decode}
&\hat{{\bf s}}_i={\bf s}_i^p+{\bf s}_{\bar{i}}^p+{\bf U}_i{\bf H}_{r,i}{\bf Q}{\bf T}{\bf W}{\bf P}{\bf n}_r+{\bf U}_i{\bf n}_i, i= 1,3,\\
&\hat{{\bf s}}_i=\left[
                                                                                                      \begin{array}{c}
                                                                                                        {\bf s}_i^p+{\bf s}_{\bar{i}}^p \\
                                                                                                        {\bf s}_{\bar{i}}^r \\
                                                                                                      \end{array}
                                                                                                    \right]
+{\bf U}_i{\bf H}_{r,i}{\bf Q}{\bf T}{\bf W}{\bf P}{\bf n}_r+{\bf U}_i{\bf n}_i, i =2,4.
\end{align}
Finally, each user can decode its desired signal after applying self-interference cancellation. The DoF tuple ${\bf d}=(5,3,3,1)$ under the antenna configuration $(M_1, M_2, M_3, M_4, N)=(6,5,4,4,9)$ is thus achievable.

From this example, we see that the main challenge lies in the design of the relay compression matrix $\bf P$ at the MAC phase in response to the asymmetric information exchange within each user pair. To tackle this challenge, we have extended the GSA principle in \cite{Liu4} to the asymmetric antenna setting as in \textit{Lemma 1}. In the next subsection, we extend the idea to the general antenna configuration and present the achievable scheme to obtain the optimal DoF region when the number of antennas at the relay falls into three different regions.

\subsection{$N \geq M_1+M_2$}
In this subsection, we present the DoF-region achievability when $N \geq M_1+M_2$. In this case, the DoF region \eqref{D} in \textit{Theorem 1} can be simplified as

\begin{subequations}\label{D_1}
\begin{align}\nonumber
{\cal D}_1^{*}=&\Big\{(d_1,d_2,d_3,d_4) \in \mathbb{R}_{+}^4:\\\label{D_1_a}
&~~~d_{1} \leq M_2 \\\label{D_1_b}
&~~~d_{2} \leq M_2 \\\label{D_1_c}
&~~~d_{3} \leq M_4 \\\label{D_1_d}
&~~~d_{4} \leq M_4 \\\label{D_1_e}
&~~~d_{1}+d_{2}+d_{3} \leq N \\\label{D_1_f}
&~~~d_{1}+d_{2}+d_{4} \leq N \\\label{D_1_g}
&~~~d_{1}+d_{3}+d_{4} \leq N \\\label{D_1_h}
&~~~d_{2}+d_{3}+d_{4} \leq N \Big\}.
\end{align}
\end{subequations}

Due to the symmetry between user $1$ and its pairing user $2$ as well as the symmetry between user $3$ and its pairing user $4$, we focus on the DoF tuple ${\bf d} \in {\cal D}_1^{*}$ where $d_1 \geq d_2$ and $d_3 \geq d_4$. Thus, besides $d_2$ (or $d_4$) pairs of independent data streams to be exchanged and aligned between user $1$ (or $3$) and user $2$ (or $4$), there are additional $d_1-d_2$ (or $d_3-d_4$) data streams to be sent from user $1$ (or $3$) to user $2$ (or $4$). We assume that user $1$ and user $3$ only utilize $M_2$ and $M_4$ antennas, respectively, in this case by antenna deactivation.

During the MAC phase, since the relay needs to decode $d_1 + d_3$ independent data streams (including both $d_2+d_4$ network-coded symbols and $d_1-d_2 + d_3-d_4 $ individual symbols), we compress the signal received at the relay by a full-rank compression matrix ${\bf P} \in {\mathbb C}^{J \times N}$, where
\begin{align}
J=d_1+d_3.
\end{align}
It is worth mentioning that $J \leq N$ is satisfied for all DoF tuples in ${\cal D}_1^{*}$ from \eqref{D_e}-\eqref{D_h}. In the $J$-dimensional compressed subspace of the relay, the first $d_2$ dimensions are used for the $d_2$ pairs of data streams transmitted from user $1$ and user $2$ to align so as to form network-coded symbols. Similarly, the second $d_4$ dimensions are used for the $d_4$ pairs of data streams from user $3$ and user $4$. The remaining $(d_1-d_2)$ and $(d_3-d_4)$ dimensions are used to decode the additional $(d_1-d_2)$ data streams sent from user $1$ to user $2$ and the additional $(d_3-d_4)$ data streams sent from user $3$ to user $4$, respectively.

According to \textit{Lemma 1}, we design ${\bf P}$, $\{{\bf V}_i^p \mid \forall i\}$, ${\bf V}_{1}^r$, and ${\bf V}_{3}^r$ such that
\begin{subequations}\label{GSA}
\begin{align}\label{GSA_1}
&{\bf P}{\bf H}_{1,r}{\bf V}_1^p={\bf P}{\bf H}_{2,r}{\bf V}_2^p,\\\label{GSA_2}
&{\bf P}{\bf H}_{3,r}{\bf V}_3^p={\bf P}{\bf H}_{4,r}{\bf V}_4^p,\\\label{GSA_3}
&\textrm{rank}([{\bf V}_{1}^p~{\bf V}_{1}^r])=d_1,\\\label{GSA_4}
&\textrm{rank}([{\bf V}_{3}^p~{\bf V}_{3}^r])=d_3.
\end{align}
\end{subequations}
Here, the definitions of the precoding matrices $\{{\bf V}_{i}^p \mid i=1,2,3,4\}$ and $\{{\bf V}_{i}^r \mid i=1,3\}$ are given in \eqref{y_r_E}. We separate the design of ${\bf P}$ and $\{{\bf V}_i^p \mid \forall i\}$ into four cases: (I) $d_1+d_2+d_3 \geq 2M_2$ and $d_1+d_3+d_4 \geq 2M_4$; (II) $d_1+d_2+d_3 \geq 2M_2$ and $d_1+d_3+d_4 < 2M_4$; (III) $d_1+d_2+d_3 < 2M_2$ and $d_1+d_3+d_4 \geq 2M_4$;  (IV) $d_1+d_2+d_3 < 2M_2$ and $d_1+d_3+d_4 < 2M_4$;

\subsubsection{Case I}
First, we consider the DoF tuples satisfying $d_1+d_2+d_3 \geq 2M_2$ and $d_1+d_3+d_4 \geq 2M_4$. We separate ${\bf P}$ into three parts as
\begin{align}
{\bf P}=\left[\begin{array}{c}
                {\bf P}_1 \\
                {\bf P}_2 \\
                {\bf P}_3
              \end{array}
\right],
\end{align}
where ${\bf P}_1$ is a $(d_1+d_2+d_3-2M_2) \times N$ submatrix, ${\bf P}_2$ is a $(d_1+d_3+d_4-2M_4) \times N$, and ${\bf P}_3$ is a $(2M_2+2M_4-d_1-d_2-d_3-d_4) \times N$ submatrix. Here, ${\bf P}_3$ exists due to the fact of ${\bf d} \in {\cal D}_1^{*}$ and \eqref{D_1_a}-\eqref{D_1_d}. We design ${\bf P}_1$ and ${\bf P}_2$ as
\begin{align}\label{I_A_P}
&\textrm{span}\left({\bf P}_1^T\right)\subseteq\textrm{null} \left(\left[{\bf H}_{1,r}~-{\bf H}_{2,r}\right]\right)^T,\\
&\textrm{span}\left({\bf P}_2^T\right)\subseteq\textrm{null} \left(\left[{\bf H}_{3,r}~-{\bf H}_{4,r}\right]\right)^T.
\end{align}
${\bf P}_3$ can be designed randomly as long as ${\bf P}$ has full row rank. Here, ${\bf P}_1$ exists because the dimension of the null space of $\left(\left[{\bf H}_{1,r}~-{\bf H}_{2,r}\right]\right)^T$ is $N-2M_2$, which is greater than or equal to $d_1+d_2+d_3-2M_2$ from the fact \eqref{D_1_e}; ${\bf P}_2$ exists because the dimension of the null space of $\left(\left[{\bf H}_{3,r}~-{\bf H}_{4,r}\right]\right)^T$ is $N-2M_4$, which is greater than or equal to $d_1+d_3+d_4-2M_4$ from the fact \eqref{D_1_g}.

Then, ${\bf P}\left[{\bf H}_{1,r}~-{\bf H}_{2,r}\right]$ can be expressed as
\begin{align}\label{I_A_PH12}
{\bf P}\left[{\bf H}_{1,r}~-{\bf H}_{2,r}\right]=\left[\begin{array}{cc}
                                                 \bf 0 & \bf 0 \\
                                                 {\bf P}_2{\bf H}_{1,r} & {\bf P}_2{\bf H}_{2,r}\\
                                                 {\bf P}_3{\bf H}_{1,r} & {\bf P}_3{\bf H}_{2,r}
                                               \end{array}
\right],
\end{align}
with rank $2M_2-d_2$. We then design the precoding matrices ${\bf V}_1^p$ and ${\bf V}_2^p$ as
\begin{align}\label{I_A_V_1}
\textrm{span}\left(\left[\begin{array}{c}
        {\bf V}_1^p \\
        {\bf V}_2^p
      \end{array}
\right] \right) \subseteq \textrm{null} \left({\bf P}\left[{\bf H}_{1,r}~-{\bf H}_{2,r}\right]\right),
\end{align}
Here, ${\bf V}_1^p$ and ${\bf V}_2^p$ exists because the dimension of the null space of ${\bf P}\left[{\bf H}_{1,r}~-{\bf H}_{2,r}\right]$ is $2M_2-(2M_2-d_2)=d_2$. Therefore, the alignment condition in \eqref{GSA_1} is satisfied.

Similarly, ${\bf P}\left[{\bf H}_{3,r}~-{\bf H}_{4,r}\right]$ can be expressed as
\begin{align}\label{I_A_PH34}
{\bf P}\left[{\bf H}_{3,r}~-{\bf H}_{4,r}\right]=\left[\begin{array}{cc}
                                                 {\bf P}_1{\bf H}_{3,r} & {\bf P}_1{\bf H}_{4,r}\\
                                                 \bf 0 & \bf 0\\
                                                 {\bf P}_3{\bf H}_{3,r} & {\bf P}_3{\bf H}_{4,r}
                                               \end{array}
\right],
\end{align}
with rank $2M_4-d_4$. We then design the precoding matrices ${\bf V}_3^p$ and ${\bf V}_4^p$ as
\begin{align}\label{I_A_V_2}
\textrm{span}\left(\left[\begin{array}{c}
        {\bf V}_3^p \\
        {\bf V}_4^p
      \end{array}
\right] \right) \subseteq \textrm{null} \left({\bf P}\left[{\bf H}_{3,r}~-{\bf H}_{4,r}\right]\right),
\end{align}
Here, ${\bf V}_3^p$ and ${\bf V}_4^p$ exists because the dimension of the null space of ${\bf P}\left[{\bf H}_{3,r}~-{\bf H}_{4,r}\right]$ is $2M_4-(2M_4-d_4)=d_4$. Thus, the alignment condition \eqref{GSA_2} is satisfied. The remaining two precoding matrices ${\bf V}_1^r$ and ${\bf V}_3^r$ can be designed randomly as long as $[{\bf V}_1^p~{\bf V}_1^r]$ and $[{\bf V}_3^p~{\bf V}_3^r]$ have full column rank, so that \eqref{GSA_3} and \eqref{GSA_4} hold. For presentation simplicity, the design of $\{{\bf V}_1^r,{\bf V}_3^r\}$ will be skipped in the remaining part of this section since the criterion is the same.

\subsubsection{Case II}
Second, we consider the DoF tuples satisfying $d_1+d_2+d_3 \geq 2M_2$ and $d_1+d_3+d_4 < 2M_4$. We separate ${\bf P}$ into two parts as
\begin{align}
{\bf P}=\left[\begin{array}{c}
                {\bf P}_1 \\
                {\bf P}_2
              \end{array}
\right],
\end{align}
where ${\bf P}_1$ is a $(d_1+d_2+d_3-2M_2) \times N$ submatrix, ${\bf P}_2$ is a $(2M_2-d_2) \times N$. We design ${\bf P}_1$ by following \eqref{I_A_P}, and design ${\bf P}_2$ randomly as long as ${\bf P}$ has full row rank. Then, ${\bf P}\left[{\bf H}_{1,r}~-{\bf H}_{2,r}\right]$ can be expressed as
\begin{align}
{\bf P}\left[{\bf H}_{1,r}~-{\bf H}_{2,r}\right]=\left[\begin{array}{cc}
                                                 \bf 0 & \bf 0 \\
                                                 {\bf P}_2{\bf H}_{1,r} & {\bf P}_2{\bf H}_{2,r}
                                               \end{array}
\right],
\end{align}
with rank $2M_2-d_2$. We design ${\bf V}_1^p$ and ${\bf V}_2^p$ according to \eqref{I_A_V_1}. Here, ${\bf V}_1^p$ and ${\bf V}_2^p$ exists because the dimension of the null space of ${\bf P}\left[{\bf H}_{1,r}~-{\bf H}_{2,r}\right]$ is $2M_2-(2M_2-d_2)=d_2$. The rank of ${\bf P}\left[{\bf H}_{3,r}~-{\bf H}_{4,r}\right]$ is $d_1+d_3$. We design ${\bf V}_3^p$ and ${\bf V}_4^p$ according to \eqref{I_A_V_2}. Here, ${\bf V}_3^p$ and ${\bf V}_4^p$ exists because the dimension of the null space of ${\bf P}\left[{\bf H}_{3,r}~-{\bf H}_{4,r}\right]$ is $2M_4-(d_1+d_3)$, which is greater than or equal to $d_4$ from the fact that $d_1+d_3+d_4 < 2M_4$.

\subsubsection{Case III}
Third, we consider the DoF tuples satisfying $d_1+d_2+d_3 < 2M_2$ and $d_1+d_3+d_4 \geq 2M_4$. This case can be converted into Case II by swapping the user indexes: $1 \leftrightarrow 3$ and $2 \leftrightarrow 4$. Then, the proof follows immediately from that of Case II.

\subsubsection{Case IV}
Finally, we consider the DoF tuples satisfying $d_1+d_2+d_3 < 2M_2$ and $d_1+d_3+d_4 < 2M_4$. This case is trivial since we can design ${\bf P}$ randomly as long as it has full row rank. The rank of ${\bf P}\left[{\bf H}_{1,r}~-{\bf H}_{2,r}\right]$ is $d_1+d_3$. We design ${\bf V}_1^p$ and ${\bf V}_2^p$ according to \eqref{I_A_V_1}. Here, ${\bf V}_1^p$ and ${\bf V}_2^p$ exists because the dimension of the null space of ${\bf P}\left[{\bf H}_{1,r}~-{\bf H}_{2,r}\right]$ is $2M_2-(d_1+d_3)$, which is greater than or equal to $d_2$ due to $d_1+d_2+d_3 < 2M_2$. The rank of ${\bf P}\left[{\bf H}_{3,r}~-{\bf H}_{4,r}\right]$ is $d_1+d_3$. We design ${\bf V}_3^p$ and ${\bf V}_4^p$ according to \eqref{I_A_V_2}. Here, ${\bf V}_3^p$ and ${\bf V}_4^p$ exists because the dimension of the null space of ${\bf P}\left[{\bf H}_{3,r}~-{\bf H}_{4,r}\right]$ is $2M_4-(d_1+d_3)$, which is greater than or equal to $d_4$ due to $d_1+d_3+d_4 < 2M_4$.

Combining Case I-IV, we have shown the design of ${\bf P}$, $\{{\bf V}_i^p\}_{i=1}^{4}$ and $\{{\bf V}_1^r,{\bf V}_3^r\}$ to meet \eqref{GSA}. The signal after compression at the relay can be expressed similarly as in \eqref{Pyr1}. Then we detect the network-coded symbols, ${\bf s}_1^p+{\bf s}_2^p$ and ${\bf s}_3^p+{\bf s}_4^p$, as well as the remaining symbols, ${\bf s}_{1}^r$ and ${\bf s}_{3}^r$, by introducing a zero-forcing matrix as in \eqref{W}.

The above precoding design directly carries over to the BC phase due to the symmetry between the MAC and the BC phases and is thus omitted. Therefore, all the DoF tuples in ${\cal D}_1^{*}$ are achievable.

\textit{Remark 4}: The above discussion can be readily generalized to a rational $J$ by using the technique of symbol extension. We refer interested readers to \cite{Wang3,Liu4} for details.

\subsection{$M_3+M_4 \leq N < M_1+M_2$}
In this section, we present the DoF-region achievability when $N < M_1+M_2$ and $N\geq M_3+M_4$. In this case, the DoF region \eqref{D} in \textit{Theorem 1} can be simplified as

\begin{subequations}\label{D_2}
\begin{align}\nonumber
{\cal D}_2^{*}=&\Big\{(d_1,d_2,d_3,d_4) \in \mathbb{R}_{+}^4:\\\label{D_2_a}
&~~~d_{1} \leq M_2 \\\label{D_2_b}
&~~~d_{2} \leq M_2 \\\label{D_2_c}
&~~~d_{3} \leq M_4 \\\label{D_2_d}
&~~~d_{4} \leq M_4 \\\label{D_2_e}
&~~~d_{1}+d_{2}+d_{3} \leq M_1+M_2 \\\label{D_2_f}
&~~~d_{1}+d_{2}+d_{4} \leq M_1+M_2 \\\label{D_2_g}
&~~~d_{1}+d_{3}+d_{4} \leq N \\\label{D_2_h}
&~~~d_{2}+d_{3}+d_{4} \leq N \Big\}.
\end{align}
\end{subequations}

Due to the symmetry between the two users in a pair, it suffices to only consider a DoF tuple ${\bf d} \in {\cal D}_2^{*}$ with $d_1 \geq d_2$ and $d_3 \geq d_4$. We assume that user $1$ only utilizes $M_2$ antennas in this case by antenna deactivation.

The basic idea is the same as that in the previous subsection. We only present the design of ${\bf P}$ and $\{{\bf V}_i^p \mid \forall i\}$ to satisfy \eqref{GSA} here. We separate the design of ${\bf P}$ and $\{{\bf V}_i^p \mid \forall i\}$ into two cases: (I) $d_1+d_3+d_4 \geq 2M_4$;  (II) $d_1+d_3+d_4 < 2M_4$.

\subsubsection{Case I}
First, we consider the DoF tuples satisfying $d_1+d_3+d_4 \geq 2M_4$. The example we illustrated in Section V-A belongs to this case. We separate ${\bf P}$ into two parts as
\begin{align}
{\bf P}=\left[\begin{array}{c}
                {\bf P}_1 \\
                {\bf P}_2
              \end{array}
\right],
\end{align}
where ${\bf P}_1$ is a $(d_1+d_3+d_4-2M_4) \times N$ submatrix, and ${\bf P}_2$ is a $(2M_4-d_4) \times N$. We design ${\bf P}_1$ as
\begin{align}\label{I_A_PP}
&\textrm{span}\left({\bf P}_1^T\right)\subseteq\textrm{null} \left(\left[{\bf H}_{3,r}~-{\bf H}_{4,r}\right]\right)^T,
\end{align}
and ${\bf P}_2$ is designed randomly as long as ${\bf P}$ has full row rank. Here, ${\bf P}_1$ exists because the dimension of the null space of $\left(\left[{\bf H}_{3,r}~-{\bf H}_{4,r}\right]\right)^T$ is $N-2M_4$, which is greater than or equal to $d_1+d_3+d_4-2M_4$ from \eqref{D_2_g}. Then, the rank of ${\bf P}\left[{\bf H}_{1,r}~-{\bf H}_{2,r}\right]$ is $d_1+d_3$. We design ${\bf V}_1^p$ and ${\bf V}_2^p$ according to \eqref{I_A_V_1}. Here, ${\bf V}_1^p$ and ${\bf V}_2^p$ exists because the dimension of the null space of ${\bf P}\left[{\bf H}_{1,r}~-{\bf H}_{2,r}\right]$ is $M_1+M_2-(d_1+d_3)$, which is greater than or equal to $d_2$ from \eqref{D_2_e}. Then, ${\bf P}\left[{\bf H}_{3,r}~-{\bf H}_{4,r}\right]$ can be expressed as
\begin{align}
{\bf P}\left[{\bf H}_{3,r}~-{\bf H}_{4,r}\right]=\left[\begin{array}{cc}
                                                 \bf 0 & \bf 0 \\
                                                 {\bf P}_2{\bf H}_{3,r} & {\bf P}_2{\bf H}_{4,r}
                                               \end{array}
\right],
\end{align}
with rank $2M_4-d_4$. We design ${\bf V}_3^p$ and ${\bf V}_4^p$ according to \eqref{I_A_V_2}. Here, ${\bf V}_3^p$ and ${\bf V}_4^p$ exists because the dimension of the null space of ${\bf P}\left[{\bf H}_{3,r}~-{\bf H}_{4,r}\right]$ is $2M_4-(2M_4-d_4)=d_4$.

\subsubsection{Case II}
Second, we consider the DoF tuples satisfying $d_1+d_3+d_4 < 2M_4$. We design ${\bf P}$ randomly, which is a full-rank matrix. The rank of ${\bf P}\left[{\bf H}_{1,r}~-{\bf H}_{2,r}\right]$ is $d_1+d_3$. We design ${\bf V}_1^p$ and ${\bf V}_2^p$ according to \eqref{I_A_V_1}. Here, ${\bf V}_1^p$ and ${\bf V}_2^p$ exists because the dimension of the null space of ${\bf P}\left[{\bf H}_{1,r}~-{\bf H}_{2,r}\right]$ is $M_1+M_2-(d_1+d_3)$, which is greater than or equal to $d_2$ from \eqref{D_2_e}. The rank of ${\bf P}\left[{\bf H}_{3,r}~-{\bf H}_{4,r}\right]$ is $d_1+d_3$. We design ${\bf V}_3^p$ and ${\bf V}_4^p$ according to \eqref{I_A_V_2}. Here, ${\bf V}_3^p$ and ${\bf V}_4^p$ exists because the dimension of the null space of ${\bf P}\left[{\bf H}_{3,r}~-{\bf H}_{4,r}\right]$ is $2M_4-(d_1+d_3)$, which is greater than or equal to $d_4$ from $d_1+d_3+d_4 < 2M_4$.

The above precoding design directly carries over to the BC phase due to the symmetry between the MAC and the BC phases and is thus omitted. Therefore, all the DoF tuples in ${\cal D}_2^{*}$ are achievable.

\subsection{Case 3: $N < M_3+M_4$}
In this section, we present the DoF-region achievability when $N < M_3+M_4$. In this case, the DoF region \eqref{D} in \textit{Theorem 1} can be simplified as

\begin{subequations}\label{D_3}
\begin{align}\nonumber
{\cal D}_3^{*}=&\Big\{(d_1,d_2,d_3,d_4) \in \mathbb{R}_{+}^4:\\\label{D_3_a}
&~~~d_{1} \leq M_2 \\\label{D_3_b}
&~~~d_{2} \leq M_2 \\\label{D_3_c}
&~~~d_{3} \leq M_4 \\\label{D_3_d}
&~~~d_{4} \leq M_4 \\\label{D_3_e}
&~~~d_{1}+d_{3} \leq N \\\label{D_3_f}
&~~~d_{1}+d_{4} \leq N \\\label{D_3_g}
&~~~d_{2}+d_{3} \leq N \\\label{D_3_h}
&~~~d_{2}+d_{4} \leq N \\\label{D_3_i}
&~~~d_{1}+d_{2}+d_{3} \leq M_1+M_2 \\\label{D_3_j}
&~~~d_{1}+d_{2}+d_{4} \leq M_1+M_2 \\\label{D_3_k}
&~~~d_{1}+d_{3}+d_{4} \leq M_3+M_4 \\\label{D_3_l}
&~~~d_{2}+d_{3}+d_{4} \leq M_3+M_4 \Big\}.
\end{align}
\end{subequations}

Due to the symmetry between the two users in a pair, it suffices to focus on a DoF tuple ${\bf d} \in {\cal D}_3^{*}$ satisfying $d_1 \geq d_2$ and $d_3 \geq d_4$.

The basic idea is the same as that in the previous subsections. We only present the design of ${\bf P}$ and $\{{\bf V}_i^p \mid \forall i\}$ to satisfy \eqref{GSA} here. Once we obtain ${\bf P}$ and $\{{\bf V}_i^p \mid \forall i\}$, then ${\bf U}_i,{\bf Q},{\bf T},{\bf W}$ can be designed as \eqref{W}, \eqref{GSA_BC} and \eqref{T}, similarly. We use the antenna deactivation method at the relay, i.e., the relay only utilize $d_1+d_3$ antennas. The rank of ${\bf P}\left[{\bf H}_{1,r}~-{\bf H}_{2,r}\right]$ is $d_1+d_3$. We design ${\bf V}_1^p$ and ${\bf V}_2^p$ according to \eqref{I_A_V_1}. Here, ${\bf V}_1^p$ and ${\bf V}_2^p$ exists because the dimension of the null space of ${\bf P}\left[{\bf H}_{1,r}~-{\bf H}_{2,r}\right]$ is $M_1+M_2-(d_1+d_3)$, which is greater than or equal to $d_2$ from \eqref{D_3_i}. The rank of ${\bf P}\left[{\bf H}_{3,r}~-{\bf H}_{4,r}\right]$ is $d_1+d_3$. We design ${\bf V}_3^p$ and ${\bf V}_4^p$ according to \eqref{I_A_V_2}. Here, ${\bf V}_3^p$ and ${\bf V}_4^p$ exists because the dimension of the null space of ${\bf P}\left[{\bf H}_{3,r}~-{\bf H}_{4,r}\right]$ is $M_3+M_4-(d_1+d_3)$, which is greater than or equal to $d_4$ from \eqref{D_3_k}.

The above precoding design directly carries over to the BC phase due to the symmetry between the MAC and the BC phases and is thus omitted. Therefore, all the DoF tuples in ${\cal D}_3^{*}$ are achievable.

\section{Sum DoF (Proof of Corollary 1)}
In this section, we prove the optimal sum DoF of the asymmetric two-pair MIMO TWRC in \textit{Corollary 1}.

Given the optimal DoF region ${\cal D}^{*}$ in \textit{Theorem 1}, the optimal sum DoF $d_{\Sigma}^{*}$ can be found by solving the following optimization problem:
\begin{align}\label{OP_1}
&d_{\Sigma}^{*}=\max_{{\bf d} \in {\cal D}^{*}}~d_1+d_2+d_3+d_4.
\end{align}
It is clear that problem \eqref{OP_1} is a linear optimization problem with 4 variables and 12 constraints. The optimal solution can be obtained numerically \cite{Boyd}. However, we are interested in finding its closed-form expression to complete the DoF analysis. The feasible region for the problem \eqref{OP_1}, i.e., the optimal DoF region ${\cal D}^{*}$ specified by \eqref{D_a}-\eqref{D_l}, is a polytope in a $4$-dimensional space. The optimal solution to $\{d_i\}_{i=1}^4$ must be located in one of the vertexes of the polytope. But it is not straightforward to find the optimal solution, as there are $\binom{12}{4}=495$ candidate vertexes\footnote{Each candidate vertex is given by letting 4 out of the 12 inequalities in \eqref{D} take equality.}. This motivates us to reduce the search space by exploiting the structural properties of the optimal solution of problem \eqref{OP_1}. To proceed, we shall present the following useful lemma.

\textit{Lemma 2}: If a DoF tuple $Q_1=(d_1,d_2,d_3,d_4)$ is an optimal solution to \eqref{OP_1}, then $Q_2=(d_1',d_2',d_3',d_4')$ is also an optimal solution to \eqref{OP_1}, where $d_1'=d_2'=\frac{d_1+d_2}{2}$ and $d_3'=d_4'=\frac{d_3+d_4}{2}$.
\begin{proof}
It is clear that the objective value for $Q_1$ and $Q_2$ are the same. Thus, it remains to show that $Q_2$ is also located in the polytope generated by ${\cal D}^{*}$. We show this in three steps.

Step 1 (Constraints \eqref{D_a}-\eqref{D_d}): Since $Q_1$ is a feasible solution to \eqref{OP_1}, we have $\max\{d_1,d_2\} \leq \min\{M_1,M_2\}$ and $\max\{d_3,d_4\} \leq \min\{M_3,M_4\}$. Then
\begin{align}\nonumber
d_1'=d_2'=\frac{d_1+d_2}{2} \leq  \max\{d_1,d_2\} \leq  \min\{M_1,M_2\}.
\end{align}
and
\begin{align}\nonumber
d_3'=d_4'=\frac{d_3+d_4}{2}\leq  \max\{d_3,d_4\}\leq  \min\{M_3,M_4\}.
\end{align}
Hence, $Q_2$ satisfies constraints \eqref{D_a}-\eqref{D_d}.

Step 2 (Constraints \eqref{D_e}-\eqref{D_h}): Since $Q_1$ is a feasible solution to \eqref{OP_1}, from \eqref{D_e}-\eqref{D_h}, we have
\begin{align}
\max\{d_1,d_2\}+\max\{d_3,d_4\} \leq N.
\end{align}
Then
\begin{align}\nonumber
&\max\{d_1',d_2'\}+\max\{d_3',d_4'\}\\\nonumber
\leq & \max\{d_1,d_2\}+\max\{d_3,d_4\}\\
\leq & N,
\end{align}
implying that $Q_2$ satisfies constraints \eqref{D_e}-\eqref{D_h}.

Step 3 (Constraints \eqref{D_i}-\eqref{D_l}): $Q_2$ satisfies constraints \eqref{D_i}-\eqref{D_l} since $d_1+d_2=d_1'+d_2'$ and $d_3+d_4=d_3'+d_4'$.

Therefore, $Q_2$ is a feasible DoF tuple and \textit{Lemma 2} is proved.
\end{proof}

\textit{Lemma 2} reveals that enforcing symmetric pairwise data exchange, i.e., $d_{i}=d_{\bar{i}}$, does not sacrifice the optimality of the sum DoF. Based on this, the optimization problem \eqref{OP_1} can be simplified as
\begin{subequations}\label{OP_2}
\begin{align}\nonumber
&\max_{\{d_2,d_4\}}~d_2+d_4\\\label{OP_2_a}
s.t.~~& d_2 \leq M_2,\\\label{OP_2_b}
& d_4 \leq M_4,\\\label{OP_2_c}
& d_2+d_4 \leq N,\\\label{OP_2_d}
& 2d_2+d_4 \leq \max\{M_1+M_2,N\},\\\label{OP_2_e}
& d_2+2d_4 \leq \max\{M_3+M_4,N\},\\\label{OP_2_f}
& d_{i} \geq 0,~~\forall i.
\end{align}
\end{subequations}
Here the problem \eqref{OP_2} only contains two variables and six constraints. It is now more tractable to search over all the vertexes of the new polytope generated by \eqref{OP_2_a}-\eqref{OP_2_f}. The optimal sum DoF and the corresponding vertexes are thus obtained and presented in \textsc{Table} \ref{vertex_case_1} $(N \geq M_1+M_2)$, \textsc{Table} \ref{vertex_case_2} $(M_3+M_4 \leq N < M_1+M_2)$ and \textsc{Table} \ref{vertex_case_3} $(N < M_3+M_4)$.

\begin{table}[]
\centering
\caption{Optimal sum DoF and their corresponding vertexes when $N \geq M_1+M_2$}
\label{vertex_case_1}
\begin{tabular}{|l|l|}
\hline
Optimal sum DoF & Achieving vertex                \\ \hline
$2M_2+2M_4$ & \begin{tabular}[c]{@{}l@{}}$d_2=M_2$\\ $d_4=M_4$\end{tabular}                 \\ \hline
$\frac{4}{3}N$ & \begin{tabular}[c]{@{}l@{}}$d_2=\frac{N}{3}$\\ $d_4=\frac{N}{3}$\end{tabular} \\ \hline
$M_2+N$ & \begin{tabular}[c]{@{}l@{}}$d_2=M_2$\\ $d_4=\frac{N-M_2}{2}$\end{tabular}     \\ \hline
$M_4+N$ & \begin{tabular}[c]{@{}l@{}}$d_2=\frac{N-M_4}{2}$\\ $d_4=M_4$\end{tabular}     \\ \hline
\end{tabular}
\end{table}

\begin{table}[]
\centering
\caption{Optimal sum DoF and their corresponding vertexes when $M_3+M_4 \leq N < M_1+M_2$}
\label{vertex_case_2}
\begin{tabular}{|l|l|}
\hline
Optimal sum DoF & Achieving vertex               \\ \hline
$2M_2+2M_4$ & \begin{tabular}[c]{@{}l@{}}$d_2=M_2$\\ $d_4=M_4$\end{tabular}                                    \\ \hline
$M_2+N$ & \begin{tabular}[c]{@{}l@{}}$d_2=M_2$\\ $d_4=\frac{N-M_2}{2}$\end{tabular}                        \\ \hline
$M_1+M_2+M_4$ & \begin{tabular}[c]{@{}l@{}}$d_2=\frac{M_1+M_2-M_4}{2}$\\ $d_4=M_4$\end{tabular}                  \\ \hline
$2N$ & \begin{tabular}[c]{@{}l@{}}$d_2=N$\\ $d_4=0$\end{tabular}                                        \\ \hline
$\frac{2(M_1+M_2+N)}{3}$ & \begin{tabular}[c]{@{}l@{}}$d_2=\frac{2M_1+2M_2-N}{3}$\\ $d_4=\frac{2N-M_1-M_2}{3}$\end{tabular} \\ \hline
\end{tabular}
\end{table}

\begin{table}[]
\centering
\caption{Optimal sum DoF and their corresponding vertexes when $N < M_3+M_4$}
\label{vertex_case_3}
\begin{tabular}{|l|l|}
\hline
Optimal sum DoF & Achieving vertex              \\ \hline
$2M_2+2M_4$ & \begin{tabular}[c]{@{}l@{}}$d_2=M_2$\\ $d_4=M_4$\end{tabular}                                                  \\ \hline
$2N$ & See Appendix A                                                                                                           \\ \hline
$M_2+M_3+M_4$ & \begin{tabular}[c]{@{}l@{}}$d_2=M_2$\\ $d_4=\frac{M_3+M_4-M_2}{2}$\end{tabular}                                    \\ \hline
$M_1+M_2+M_4$ & \begin{tabular}[c]{@{}l@{}}$d_2=\frac{M_1+M-2-M_4}{2}$\\ $d_4=M_4$\end{tabular}                                \\ \hline
$\frac{2(M_1+M_2+M_3+M_4)}{3}\}$ & \begin{tabular}[c]{@{}l@{}}$d_2=\frac{2M_1+2M_2-M_3-M_4}{3}$\\ $d_2=\frac{-M_1-M-2+2M_3+2M_4}{3}$\end{tabular} \\ \hline
\end{tabular}
\end{table}

\section{Conclusion}
In this work, we have presented a complete characterization of the optimal DoF region of the asymmetric two-pair MIMO TWRC. The proposed transmission scheme takes into account the asymmetric data exchange within each user pair and designs the relay compression matrix and all the source precoding matrices jointly using the generalized signal alignment principle. We have also derived the optimal sum DoF of the asymmetric two-pair MIMO TWRC. Our results reveal that in the asymmetric antenna setting, some antennas at certain source nodes are redundant and do not contribute to enlarge the DoF region. Our results also reveal that enforcing symmetric data exchange within each user pair does not lose the optimality of the sum DoF.

For the multi-pair MIMO TWRC with more than 2 pairs, the optimal sum DoF is still unknown even for symmetric antenna setting. Thus, determining the optimal DoF region for the multi-pair MIMO TWRC still remains open.

\section*{Appendix A}
Here, we present the vertex that achieves the optimal sum DoF $2N$ in 8 cases. Define
\begin{align}
\alpha=\min\left\{2M_2+2M_4, M_2+M_3+M_4, M_1+M_2+M_4, \frac{2(M_1+M_2+M_3+M_4)}{3}\right\}.
\end{align}
\begin{itemize}
  \item If $2M_2+2M_4=\alpha$ and $N=M_2+M_4$, then the achieving vertex is $(d_2,d_4)=(M_2, M_4)$.
  \item If $2M_2+2M_4=\alpha$ and $N<M_2+M_4$, then antenna deactivation is applied at user $2$ and user $4$ in order to set $N=M_2^u+M_4^u$, where $M_2^u$ and $M_4^u$ are respectively the numbers of antennas utilized at users $2$ and $4$ after antenna deactivation. The achieving vertex is $(d_2,d_4)=(M_2^u, M_4^u)$.
  \item If $M_2+M_3+M_4=\alpha$ and $N=\frac{M_2+M_3+M_4}{2}$, then the achieving vertex is $(d_2,d_4)=(M_2, \frac{M_3+M_4-M_2}{2})$.
  \item If $M_2+M_3+M_4=\alpha$ and $N<\frac{M_2+M_3+M_4}{2}$, then antenna deactivation is applied at user $2$, user $3$, and user $4$ in order to set $N=\frac{M_2^u+M_3^u+M_4^u}{2}$. The achieving vertex is $(d_2,d_4)=(M_2^u, \frac{M_3^u+M_4^u-M_2^u}{2})$.
  \item If $M_1+M_2+M_4=\alpha$ and $N=\frac{M_1+M_2+M_4}{2}$, then the achieving vertex is $(d_2,d_4)=(\frac{M_1+M_2-M_4}{2}, M_4)$.
  \item If $M_1+M_2+M_4=\alpha$, then antenna deactivation is applied at user $1$, user $2$, and user $4$ to ensure $N=\frac{M_1^u+M_2^u+M_4^u}{2}$. The achieving vertex is $(d_2,d_4)=(\frac{M_1^u+M_2^u-M_4^u}{2}, M_4)$.
  \item If $\frac{2(M_1+M_2+M_3+M_4)}{3}=\alpha$ and $N=\frac{2(M_1+M_2+M_3+M_4)}{3}$, then the achieving vertex is $(d_2,d_4)=(\frac{2M_1+2M_2-M_3-M_4}{3}, \frac{-M_1-M_2+2M_3+2M_4}{3})$.
  \item If $\frac{2(M_1+M_2+M_3+M_4)}{3}=\alpha$ and $N<\frac{2(M_1+M_2+M_3+M_4)}{3}$, then antenna deactivation is applied at user $1$, user $2$, user $3$, and user $4$ to ensure $N=\frac{2(M_1^u+M_2^u+M_3^u+M_4^u)}{3}$. The achieving vertex is $(d_2,d_4)=(\frac{2M_1^u+2M_2^u-M_3^u-M_4^u}{3}, \frac{-M_1^u-M_2^u+2M_3^u+2M_4^u}{3})$.
\end{itemize}

\bibliographystyle{IEEEtran}
\bibliography{IEEEabrv,reference}

\end{document}